\documentclass[sigconf]{acmart}
\renewcommand\footnotetextcopyrightpermission[1]{} 
\usepackage{amsmath,bm}
\usepackage{booktabs} 
\usepackage{url}
\usepackage{mathrsfs}
\usepackage[mathscr]{eucal}
\usepackage[linesnumbered,ruled,vlined]{algorithm2e}
\usepackage{multirow}
\usepackage{subcaption}
\usepackage{amsthm}
\usepackage{amsmath}
\usepackage{flushend}
\usepackage{wrapfig}

\usepackage{tikz}
\usetikzlibrary{arrows,positioning,automata,calc,shapes}
\usepackage{pgfplots}
\pgfplotsset{compat=newest, scaled z ticks=false} 
\pgfplotsset{plot coordinates/math parser=false}
\newlength\figureheight 
 \newlength\figurewidth

\newcommand{\squishlist}{
    \begin{list}{$\bullet$}
    { \setlength{\itemsep}{0pt}
        \setlength{\parsep}{1pt}
        \setlength{\topsep}{1pt}
        \setlength{\partopsep}{0pt}
        \setlength{\leftmargin}{1em} 
        \setlength{\labelwidth}{1em}
        \setlength{\labelsep}{0.5em}
    						 } }

\newcommand{\squishlisttwo}{
    \begin{list}{$\bullet$}
        { \setlength{\itemsep}{0pt}
            \setlength{\parsep}{0pt}
            \setlength{\topsep}{0pt}
            \setlength{\partopsep}{0pt}
            \setlength{\leftmargin}{2em}
            \setlength{\labelwidth}{1.5em}
            \setlength{\labelsep}{0.5em} } }

\newcommand{\squishend}{
    \end{list}  }
    
\setcopyright{none}
\begin{document}

\acmDOI{1111}

\acmISBN{1111}

\acmConference[111]{ACM conference}{Jan 2018}{College Station, Texas USA}
\acmYear{2018}
\copyrightyear{2018}
\acmArticle{1}
\acmPrice{00.00}

\title{Fairness-Aware Recommendation of Information Curators}

 \author{Ziwei Zhu, Jianling Wang, Yin Zhang, and James Caverlee}
 \affiliation{%
   \institution{Texas A\&M University}
 }
 \email{{zhuziwei,jwang713,zhan13679,caverlee}@tamu.edu}


\fancyhead{}

\begin{abstract}
This paper highlights our ongoing efforts to create effective information curator recommendation models that can be personalized for individual users, while maintaining important fairness properties. Concretely, we introduce the problem of information curator recommendation, provide a high-level overview of a fairness-aware recommender, and introduce some preliminary experimental evidence over a real-world Twitter dataset. We conclude with some thoughts on future directions.



\end{abstract}

\maketitle

\section{Introduction}
Information curators serve as conduits to high-quality curated content, providing unique specialized expertise, trustworthiness in decision-making, and access to novel content. For example, in the aftermath of a natural disaster, critical citizen responders on Facebook can filter through the noise to curate high-quality, informative posts, while avoiding likely misinformation \cite{maxwell2012crisees,starbird2011voluntweeters}. During breaking news, knowledgeable locals can provide access to reputable reporting and contextual insights into the developing situation \cite{dailey2014journalists,kav2016}. And in a longer-term perspective, information curators can provide deep dives into health  claims (e.g., by comparing and contrasting multiple analyses of the health benefits of new diets), products to buy (e.g., through rigorous evaluation and comparison), and insights into local governance issues (e.g., through curating analyses of proposed state amendments or local bond issues). 

Successfully  uncovering  such information curators from the massive scale of the web and social media, reliably connecting users to the appropriate curators, and ensuring fairness-preserving properties of such curators is vitally important to prompt a trustworthy information diet, to improve the quality of user experience, and to support an informed populace. In practice, most users access information curators today via a centralized platform -- like Facebook, Google, or NYTimes -- meaning that the (hidden) algorithms connecting users to curators are essentially blackboxes. As a result, users have limited understanding of what factors impact what content is surfaced to them and whether or not any bias is shaping their information diet. In our context, such bias could lead to the suppression of curators by gender, race, religious beliefs, political stances, or other factors. 


Yet there is a critical research gap in fairness-aware personalized recommendation of information curators at scale: First, many existing recommender systems focus on specific items -- like movies, songs, or books as the basis of recommendation -- rather than on information curators who organize a heterogeneous mix of high-quality curated items. And for those approaches that aim to uncover knowledgeable users in online systems -- e.g., \cite{balog2006formal, cheng2014barbecue, ghosh2012cognos, weng2010twitterrank, zhang2007expertise} -- most typically do so without an emphasis on personalized recommendation. Indeed, there is a research gap in our understanding of both (i) identifying high-quality information curators who are gateways to curated content, and not just experts; and (ii) identifying personally-relevant curators, and not just highly-rated or popular ones.  Second, there are typically complex and dynamic relationships between users, candidate curators, and topics of interest that manifest differently in heterogeneous environments. How to model such heterogeneity is critically important. And since user interests and information needs are inherently dynamic -- that is, during a crisis, our preferences may be fundamentally different from our preferences in the long-term -- there is a need for adaptable methods to handle these complex inter-relationships. Finally, as we have mentioned, most current access to information curators is mediated by centralized platforms (like search engines, social networks, and traditional news media), meaning that  personal preferences may not align with the goals of these platforms, leading to potentially biased (or even limited) access to curators. 

Toward tackling these challenges, we have begun a broad research effort to create new personalized recommendation frameworks for connecting users with high-quality information curators, even in extremely sparse, heterogeneous, and fairness-aware environments.\footnote{\url{http://faculty.cse.tamu.edu/caverlee/curators.html}} In the rest of this paper, we focus on our first steps at building a fairness-aware recommender in this context. We provide a high-level overview of the approach (more details can be found in a companion piece \cite{zhu2018cikm}), and we introduce some preliminary experimental evidence over a real-world Twitter dataset. Our hope is this can spark a discussion about the important fairness challenges facing information curator recommendation.

\begin{figure}[t!]
\centering
\includegraphics[ width=0.43\textwidth ]{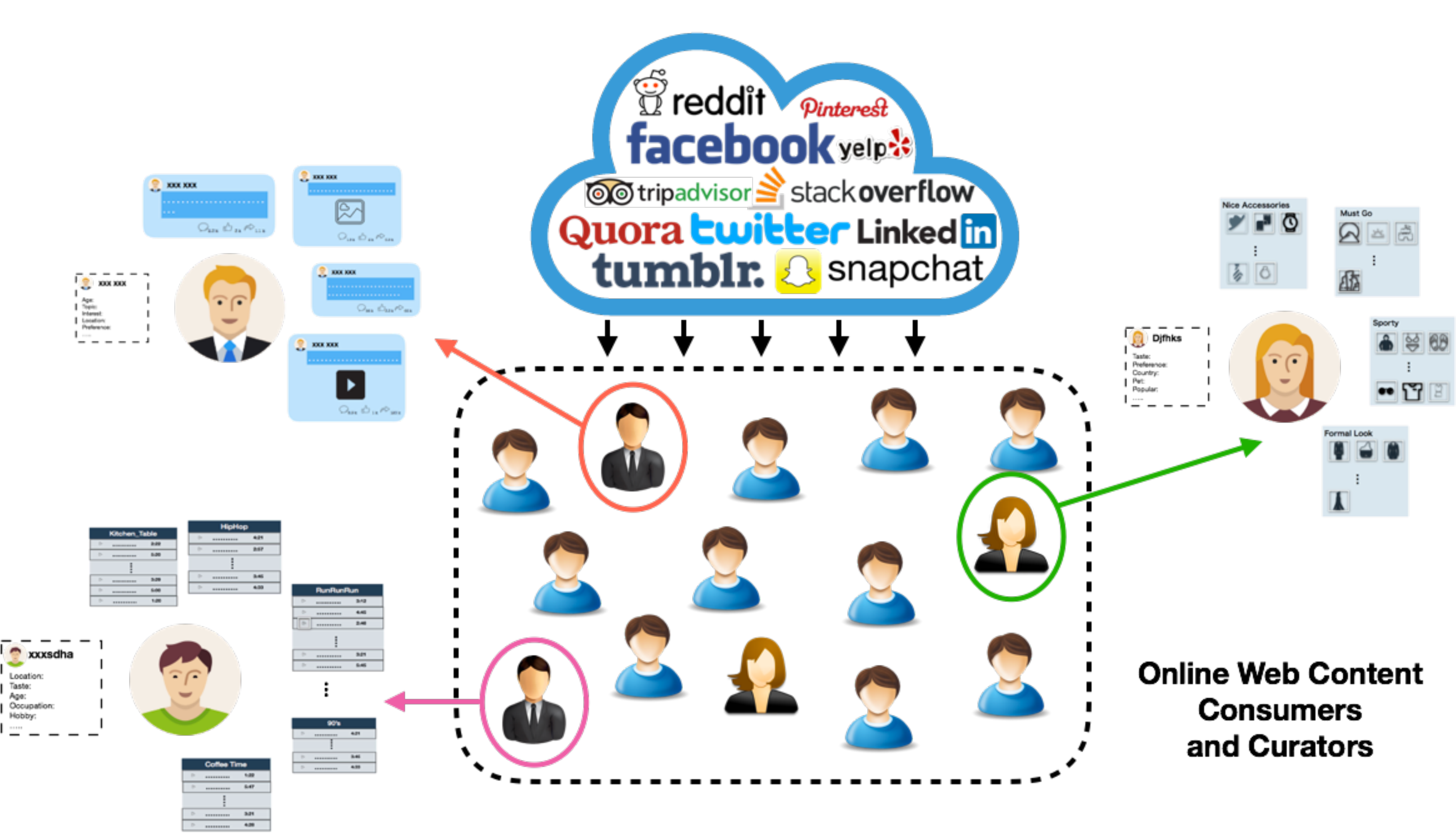} 
\caption{Information curators vary across web and social media platforms. We view curators as both high-level entities and by the curated items themselves.}
\vspace{-.3cm}
 \label{fig:example-curators}  
\end{figure}






\section{Problem Setting}
\label{sec:objectives}
We assume there exists a set of users $U=\{u_1, u_2, ..., u_N\}$, where $N$ is the total number of users. From this set $U$, there are a number of information curators denoted as $C=\{c_1, c_2 ..., c_M\}$, where $M$ is the total number of information curators. We then define the problem of \textit{personalized information curator recommendation} as: Given a user $u_i$, identify the top-$n$ personally relevant curators to $u_i$.  In practice, these information curators vary greatly. As illustrated in Figure~\ref{fig:example-curators}, these information  curators can be viewed as high-level entities (e.g., a 30-something in San Francisco with interests in entrepreneurship, a journalist in NYC focused on local governance issues) as well as by the curated items themselves (e.g., a series of blog posts analyzing a recent election, a list of product reviews). This variety and the resultant heterogeneity -- in terms of content types, social relations, motivations of curators, etc. -- place great demands on effective personalization. Further, existing expressed preferences for these curators in the forms of likes, following relationships, and other interactions are often sparse. Hence, a key challenge for personalized curator recommendation is tackling  sparsity while carefully modeling curators in  complex, noisy, and heterogeneous environments.


\section{Fairness-Aware Learning for Information Curators}  
Since user preferences for curators may be impacted by many contextual factors, we propose to  directly incorporate the multiple and varied relationships among users, curators, topics, and other factors directly into a \textit{tensor-based approach}. Such a three-dimensional tensor can be naturally extended to capture higher-order factors (e.g., by adding extra dimensions for location, reputation, and so on). Of course, we can also explore traditional matrix-based approaches in comparison with these tensor-based ones. 

As a first step, we can tackle the personalized curator recommendation problem with a basic recommendation framework using tensor factorization. Let $\bm{U}^{(1)} \in \mathbb{R}^{N \times R}$,  $\bm{U}^{(2)} \in \mathbb{R}^{M \times R}$ and $\bm{U}^{(3)} \in \mathbb{R}^{K \times R}$ be latent factor matrices for users, curators, and topics, respectively. 
The basic tensor-based curator recommendation model can be defined as:
\begin{equation*}
	\begin{aligned}
		& \underset{\bm{U}^{(n)},\bm{\mathscr{X}}}{\text{minimize}}
		& & \frac{1}{2}\|\bm{\mathscr{X}}-[\![\bm{U}^{(1)},\bm{U}^{(2)},\bm{U}^{(3)}]\!]\|^2_F + \frac{\lambda}{2}\sum_{n=1}^{3}\|\bm{U}^{(n)}\|_{F}^{2},\\
		& \text{subject to}
		& & \bm{\Psi}*\bm{\mathscr{X}}=\bm{\mathscr{T}} 
	\end{aligned}
	\label{sec3:BasicObjFun}
\end{equation*}


This basic model estimates $\bm{\mathcal{\hat{X}}}$ that approximates the original rating tensor (unknown) $\bm{\mathscr{X}}$ via learning optimal latent factor matrices $\{\bm{U}^{(n)}, n=1,2,3\}$. For each user and topic of interest, this model can recommend a ranked list of personalized curators.

\subsection{Isolating Sensitive Information}

\begin{figure}[t!]
\centering
\includegraphics[ width=0.43\textwidth ]{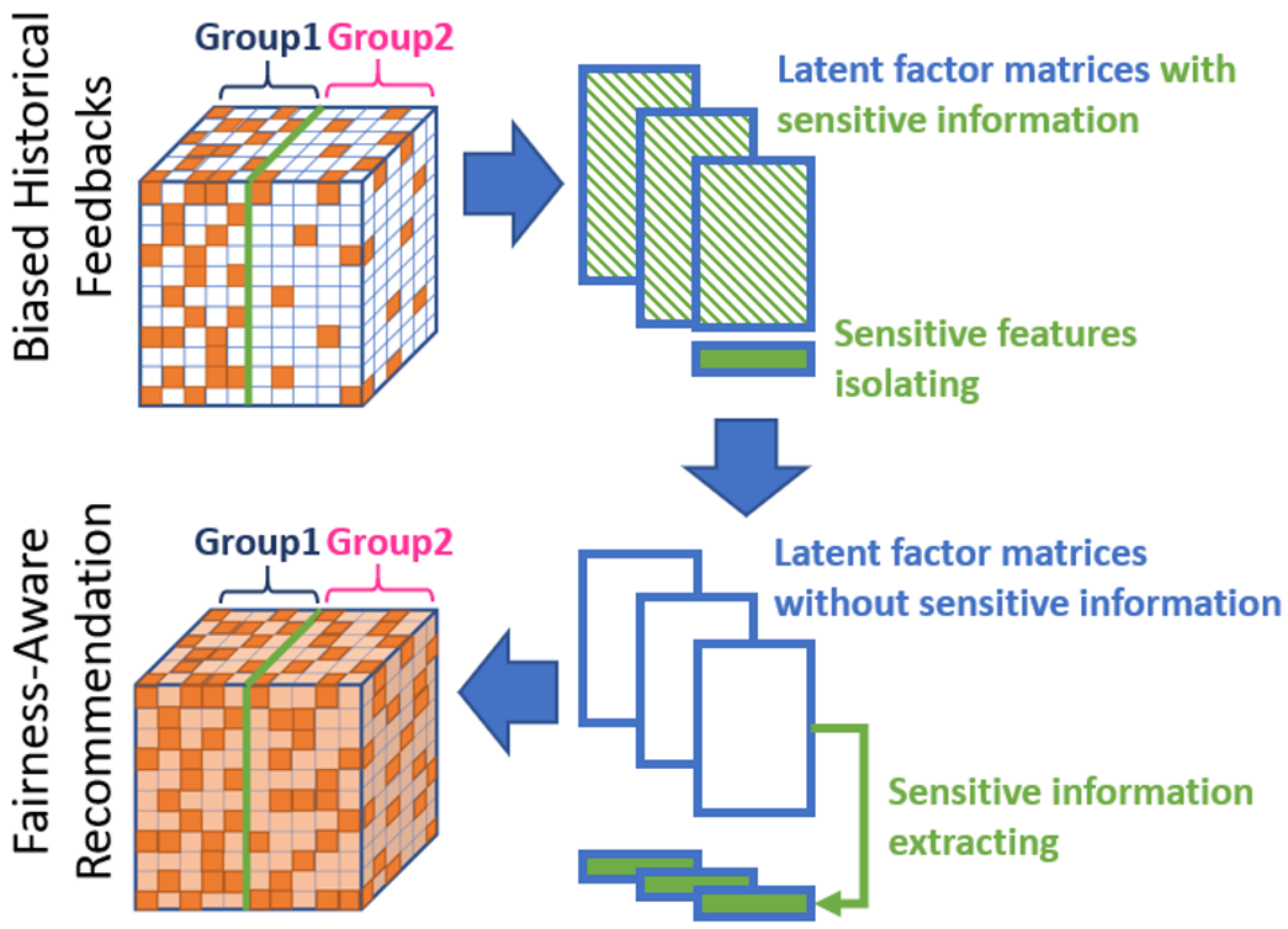} 
\caption{Overview: sensitive features are isolated (top right), then sensitive information is extracted (bottom right), resulting in fairness-aware recommendation.}
 \label{fig:Fairness} 
\vspace{-0.3cm} 
\end{figure}

However, such basic recommenders may inherit bias from the training data used to optimize them and from mis-alignment between platform goals and personal preferences. Hence, we aim to build new fairness-aware algorithms that can empower users by enhancing diversity of topics, curators, and viewpoints. As illustrated in Figure~\ref{fig:Fairness}, we aim to augment our existing methods by isolating sensitive features through a new sensitive latent factor matrix, creating a sensitive information regularizer that extracts sensitive information which can taint other latent factors, and producing fairness-enhanced recommendation by the new latent factor matrices without sensitive information. For a fuller treatment of this approach, please refer \cite{zhu2018cikm}.

For clarity, assume we have a tensor-based recommender. Such a tensor-based approach has no notion of fairness. Here, we assume that there exists a \textit{sensitive attribute} for one mode of the tensor, and this mode is a \textit{sensitive mode}. For example, the sensitive attribute could correspond to gender, age, ethnicity, location, or other domain-specific attributes of users, curators or topics in the recommenders. The feature vectors of the sensitive attributes are called the \textit{sensitive features}. Further, we call all the information related to the sensitive attributes as \textit{sensitive information}, and note that attributes other than the sensitive attributes can also contain sensitive information ~\cite{kamishima2011fairness,zemel2013learning}. While there are emerging debates about what constitutes algorithmic fairness~\cite{corbett2017algorithmic}, we adopt the commonly used notion of \textit{statistical parity}. Statistical parity encourages a recommender to ensure similar probability distributions for both the dominant group and the protected group as defined by the sensitive attributes. Formally, we denote the sensitive attribute as a random variable $S$, and the preference rating in the recommender system as a random variable $R$. Then we can formulate fairness as $\mathbf{P}[R]=\mathbf{P}[R|S]$, i.e. the preference rating is independent of the sensitive attribute. This statistical parity means that the recommendation result should be independent to the sensitive attributes. For example, a job recommender should recommend similar jobs to men and women with similar profiles. Note that some recent works~\cite{hardt2016equality,yao2017beyond,yao2017new} have argued that statistical parity may be overly strict, resulting in poor utility to end users. Our work here aims to achieve comparable utility to non-fair approaches, while providing stronger fairness.

Given this (admittedly limited) notion of fairness, the intuition of the proposed framework is that the latent factor matrices of the tensor completion model contain latent information related to the sensitive attributes, which introduces the unfairness. Therefore, by \textit{isolating} and then \textit{extracting} the sensitive information from the latent factor matrices, we may be able to improve the fairness of the recommender itself. We propose to first isolate the impact of the sensitive attribute by plugging the sensitive features into the latent factor matrix. For instance, in our user-curator-topic example where we want to enhance the recommendation fairness for curators of both genders, we can create one vector $\mathbf{s}_0$ with $1$ representing male curator and $0$ representing female curator, and another vector $\mathbf{s}_1$ with $1$ indicating female and $0$ indicating male. $\mathbf{s}_0$ and $\mathbf{s}_1$ together form a matrix, denoted as \textit{Sensitive Features} $\mathbf{S}$. We put $\mathbf{S}$ to the last two columns of the latent factor matrix of sensitive mode (the mode for curators). In this way, we construct a new \textit{sensitive latent factor matrix}, and we call the last two columns as \textit{sensitive dimensions} and the others \textit{non-sensitive dimensions}.  By isolating the sensitive features, we provide a first step toward improving the fairness of the recommender. But there may still be sensitive information that resides in non-sensitive dimensions. To extract this remaining sensitive information, we propose an additional constraint that the non-sensitive dimensions should be orthogonal to the sensitive dimensions in the sensitive latent factor matrix. After the above two steps, we can get the new latent factor matrices, whose sensitive dimensions hold features exclusively related to the sensitive attributes. And their non-sensitive dimensions are decoupled from the sensitive attributes. Thus, we can derive the fairness-enhanced recommendation by reconstructing the rating tensor only by the non-sensitive dimensions. 


\section{Experiments}
To test such an approach, we adopt a Twitter dataset introduced in ~\cite{ge2016taper} that has $589$ users, $252$ curators, and $10$ topics (e.g., news, sports). There are $16,867$ links from users to curators across these topics capturing that a user is interested in a particular curator. The sparsity of this dataset is $1.136\%$. We consider race as a sensitive attribute and aim to divide curators into two groups: whites and non-whites. We apply the Face++ (https://www.faceplusplus.com/) API to the images of each curator in the dataset to derive ethnicity. In total, we have 126 whites and 126 non-whites, with 11,612 positive ratings for white curators but only 5,255 for non-whites. Since this implicit feedback scenario has no negative observations, we randomly pick unobserved data samples to be negative feedback with probability of 0.113\% (one tenth of the sparsity). We randomly split the dataset into 70\% training and 30\% testing.

\subsection{Metrics} We consider metrics to capture recommendation quality, recommendation fairness, and the impact of eliminating sensitive information.

\medskip
\noindent\textbf{Recommendation Quality.} To measure \textit{recommendation quality}, we adopt \textbf{Precision@k} and \textbf{Recall@k}, which are defined as: 

\begin{equation*}
\begin{aligned}
Precision@k = \frac{1}{|\mathbb{U}|}\sum_{u\in \mathbb{U}}\frac{|\mathbb{O}_{u}^k \cap \mathbb{O}_{u}^+|}{k},
\end{aligned}
\end{equation*}
\begin{equation*}
\begin{aligned}
Recall@k = \frac{1}{|\mathbb{U}|}\sum_{u\in \mathbb{U}}\frac{|\mathbb{O}_{u}^k \cap \mathbb{O}_{u}^+|}{\mathbb{O}_{u}^+},
\end{aligned}
\end{equation*}
where $\mathbb{O}_{u}^+$ is the set of items user $u$ gives positive feedback to in test set and $\mathbb{O}_{u}^k$ is the predicted top-k recommended items.  
We also consider \textbf{F1@k} score, which can be calculated by $F1@k=2\cdot (Precision@k\times Recall@k)/(Precision@k + Recall@k)$. We set $k=15$ in our experiments.

\medskip
\noindent\textbf{Recommendation Fairness.} To measure \textit{recommendation fairness}, we consider two complementary metrics. The first one is the absolute difference between the mean ratings of different groups (\textbf{MAD}): 
\begin{equation*}
\begin{aligned}
\centering
MAD = \vert\dfrac{\sum R^{(0)}}{\vert R^{(0)} \vert}-\dfrac{\sum R^{(1)}}{\vert R^{(1)} \vert}\vert,
\end{aligned}
\label{equ:MAD}
\end{equation*}
where $R^{(0)}$ and $R^{(1)}$ are the predicted ratings for the two groups and $\vert R^{(i)} \vert$ is the total number of ratings for group $i$. Larger values indicate greater differences between the groups, which we interpret as unfairness.

The second fairness measure is the Kolmogorov-Smirnov statistic (\textbf{KS}), which is a nonparametric test for the equality of two distributions. The KS statistic is defined as the area difference between two empirical cumulative distributions of the predicted ratings for groups: 
\begin{equation*}
\begin{aligned}
\centering
KS = \vert\sum_{i=1}^T l\times\dfrac{\pmb{\mathscr{G}}(R^{(0)}, i)}{\vert R^{(0)} \vert}-\sum_{i=1}^T l\times\dfrac{\pmb{\mathscr{G}}(R^{(1)}, i)}{\vert R^{(1)} \vert}\vert,
\end{aligned}
\label{equ:KS}
\end{equation*}
where $T$ is the number of intervals for the empirical cumulative distribution, $l$ is the size of each interval, $\pmb{\mathscr{G}}(R^{(0)}, i)$ counts how many ratings are inside the $i^{th}$ interval for group $0$. In our experiments, we set $T=50$. Lower values of KS indicate the distributions are more alike, which we interpret as being more fair.


Note that we measure the fairness in terms of MAD and KS metrics across groups rather than within individuals, since absolute fairness for every individual may be overly strict and in opposition to personalization needs of real-world recommenders. 

\medskip





\subsection{Baselines} 



We adopt a modified Gradient Descent algorithm to optimize the proposed model and name the framework \textbf{FT} -- in comparison with two tensor-based alternatives:

\medskip
\noindent\textit{Ordinary Tensor Completion (\textbf{OTC})}: The first is the conventional CP-based tensor completion method using ALS optimization algorithm. 
This baseline incorporates no notion of fairness, so it will provide a good sense of the state-of-the-art recommendation quality we can achieve. 


\medskip
\noindent\textit{Regularization-based Tensor Completion (\textbf{RTC})}: The second one is an extension from the fairness-enhanced matrix completion with regularization method introduced in \cite{kamishima2017considerations,kamishima2013efficiency,yao2017beyond}, which adds a bias penalization term to the matrix factorization objective function. For tensor-based recommenders, we can also add a regularization term to enforce statistical parity.
We use Gradient Descent to solve this optimization problem.

\medskip
We also consider purely matrix-based approaches, which compute user preferences on curators for each topic independently. We consider matrix versions of our tensor based methods (named \textbf{FM}) corresponding to FT versus matrix baselines of \textit{Ordinary Matrix Completion (\textbf{OMC})} corresponding to OTC and \textit{Regularization-based Matrix Completion (\textbf{RMC})} corresponding to RTC.

\subsection{Experimental Results}
\label{sec:Recommending_Experts_Single}



\begin{table}[t!]
  \begin{center}\small
    \begin{tabular}[0.1\textwidth]{@{}c|cccccc@{}} 
    \hline \hline
    Methods & R@15   & P@15   & KS     & MAD      \\ \hline \hline
    OMC     & 0.4128 & 0.0942 & 0.1625 & 0.0127  \\
    OTC     & \textbf{0.4384} & \textbf{0.0958} & 0.3662 & 0.0333  \\ \hline

	RMC     & 0.1609 & 0.0702 & 0.1521 & 0.0086  \\
    RTC     & 0.3003 & 0.0515 & 0.2003 & 0.0171  \\ \hline

    FM & 0.4045 & 0.0891 & 0.0523 & 0.0037  \\
	FT & 0.4180 & 0.0870 & \textbf{0.0195} & \textbf{0.0024}  \\ \hline \hline
    \end{tabular}
  \end{center}
   \vspace{1pt}
        \caption{Comparison for recommending Twitter curators.}
\label{table:EXP_B} \vspace{-.9cm}
\end{table}

\begin{figure*}[t!]
\centering
\includegraphics[width=0.9\textwidth]{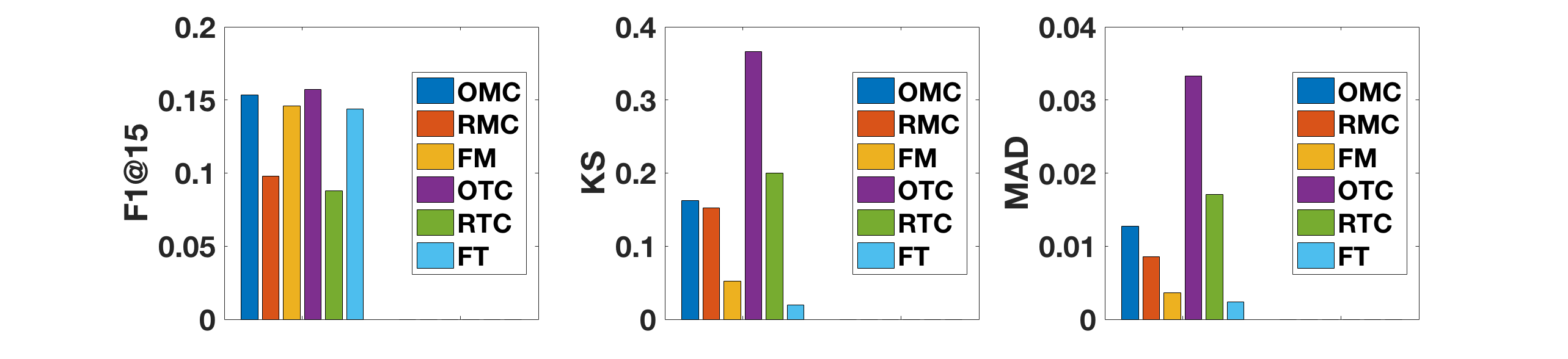} 
\vspace{-0.5cm} 
\caption{F1@15, KS, and MAD scores for all six models. The fairness-aware approaches (FM for matrix, FT for tensor) provide comparable quality (F1@15) relative to non-fair methods while providing much lower KS and MAD scores (which are proxies for statistical parity).}
 \label{fig:experiment-result} \vspace{-.3cm} 
\end{figure*}

We set 20 as the latent dimension for all the methods and fine tune all other parameters. The experiments are run three times and the averaged results are reported.

First, let's focus on the differences between matrix and tensor approaches as shown in Table ~\ref{table:EXP_B}. We observe that the tensor-based approaches mostly provide better recommendation quality (Precision@k and Recall@k) in comparison with the matrix-based approaches. Since the Twitter curation dataset is naturally multi-aspect, the tensor approaches better model the multi-way relationships among users, curators, and topics. We see that the fairness quality (KS and MAD) of matrix-based methods are better than tensor-based ones for the baselines methods (OMC vs OTC, and RMC vs RTC), but the fairness improves for our proposed methods when we move from matrix to tensor (FM vs FT). 

Second, let's consider the empirical results across approaches as present in Figure~\ref{fig:experiment-result}. We see that: (i) the proposed methods are slightly worse than OTC from the perspective of recommendation quality, but keep the difference small, and FM methods also have comparable recommendation performance with OMC; and (ii) FT provides the best fairness enhancement results, and FM also alleviates the unfairness a lot compared with other matrix-based methods. RTC and RMC improve the fairness as well, but their effects are not as good as the proposed methods.

These results suggest the potential of such a framework toward building fairness-aware information curator recommendation.%

\section{Related Work}
Many previous works  have focused on finding experts in many domains (e.g., enterprise corporate, email networks \cite{balog2006formal,campbell2003expertise,guy2013mining,liu_cikm2005,mcdonald2000expertise,serdyukov2008modeling,zhang2007expertise}), with a recent emphasis on social media \cite{ghosh2012cognos,pal_wsdm2011,weng2010twitterrank}. Building on these efforts, we have prototyped expert recommenders on Twitter \cite{cheng2014barbecue,ge2016taper,lu2015exploiting} that provide a firm foundation for the proposed research tasks here. However, there is a research gap in identifying personalized, high-quality information curators who are gateways to curated content, and not just experts or popular users.  In a related direction, many works have focused on recommendation, in which user preferences may be projected into a lower dimensional embedding space \cite{koren2009matrix, lu2015exploiting, yang2014tag, yu2012scalable}. MF and BPR-based approaches have shown good success in implicit feedback scenarios \cite{chen2012collaborative,rendle2009bpr,zhang2013combining} as in our case, and in social media scenarios \cite{Krohn-Grimberghe:2012,zhang2013combining}. In recent years, tensor factorization models are becoming popular and successfully applied in recommendation, including \cite{bhargava2015and,karatzoglou2010multiverse,hidasi2012fast}. In contrast, this project explores personalized curator recommendation by integrating multiple, heterogeneous contexts into a unified model. 

Fairness, accountability, and transparency are critical issues. Friedman \cite{friedman1996bias} defined that a computer system is biased \textit{``if it systematically and unfairly discriminate[s] against certain individuals or groups of individuals in favor of others.''} For information curator recommenders, algorithmic bias could lead to the suppression of curators by gender, race, religious beliefs, political stances, or other factors. Such algorithmic bias encodes unwanted biases in the daily experiences of millions of users, and potentially violates discrimination laws. Indeed, researchers across communities have begun actively investigating evidence of bias in existing systems and in methods towards revealing and/or overcoming this bias \cite{farahat2012effective, kulshrestha2017quantifying,pedreshi2008discrimination,zehlike2017fa,zemel2013learning}. In the recommender space, recent work has led to notions of group fairness and individual fairness in recommendation \cite{zemel2013learning}, and fairness-criteria for top-k ranking recommendation \cite{zehlike2017fa}. Kamishima summarized that recommender systems should be in adherence to laws and regulations, should be fair to all the content providers, and should exclude the influence of unwanted information \cite{kamishima2017considerations}. 

\section{Conclusion and Future Work}
This paper highlights our initial efforts at building information curator recommenders that incorporate a simple notion of fairness. In our continuing work, we are interested in generalizing our framework to consider alternative notions of fairness beyond statistical parity. By extending our framework in this direction, we can provide a more customizable approach for defining and deploying fairness-aware methods. We are also interested in exploring how to incorporate real-valued features into the framework for recommenders with explicit ratings, and in running user studies on the perceived change of fairness for our methods.

\begin{acks}
This work is, in part, supported by DARPA (\#W$911$NF-$16$-$1$-$0565$) and NSF (\#IIS-$1841138$). The views, opinions, and/or findings expressed are those of the author(s) and should not be interpreted as representing the official views or policies of the Department of Defense or the U.S. Government.
\end{acks}

\bibliographystyle{ACM-Reference-Format}
\bibliography{sample-bibliography}

\end{document}